\documentstyle[11pt,fleqn,psfig]{article}
\if@twoside \oddsidemargin 44pt \evensidemargin 82pt \marginparwidth 107pt
\else \oddsidemargin -5pt \evensidemargin -5pt
\fi
\def\etal{{\it et\thinspace al.}\ }
\begin{document}
\newcommand{\be}{\begin{equation}}
\newcommand{\ee}{\end{equation}}
\begin{center}
{\bf \Large THE IRON PROJECT}\\[5mm]
{\large Anil K. Pradhan\\Department of Astronomy, The Ohio State 
University,\\ Columbus, Ohio, USA 43210}
\end{center}
\begin{quotation}
\noindent {\bf Abstract}. Recent advances in theoretical atomic physics
have enabled large-scale calculation of atomic parameters for a variety
of atomic processes with high degree of precision. The development and 
application of these methods is the aim of the Iron Project. At present 
the primary focus is on collisional processes for all ions of iron, Fe~I~--~FeXXVI, and other iron-peak elements; new work on radiative processes has also 
been initiated.
Varied applications of the Iron Project work to X-ray astronomy are discussed,
and more general applications to other spectral ranges are pointed out. The
IP work forms the basis for more specialized projects such as the RMaX Project,
and the work on photoionization/recombination, and aims to provide a 
comprehensive and self-consistent set of accurate collsional and radiative 
cross sections, and transition probabilities, within the framework of 
relativistic close coupling formulation using the Breit-Pauli R-Matrix method.
An illustrative example is presented of how the IP data may be utilised
in the formation of X-ray spectra of the K$\alpha$ complex at 6.7 keV
from He-like Fe~XXV.

\end{quotation}

\section{Introduction}

 The main purpose of the Iron Project (IP; Hummer \etal 1993) is the continuing development
of relativistic methods for the calculations of atomic data for
electron impact excitation and radiative transitions in iron and
iron-peak elements. Its forerunner, the Opacity Project (OP; Seaton
\etal 1994; The Opacity Project Team 1995), was concerned with the calculation of radiative parameters for
astrophysically abundant elements,
oscillator strengths and photoionization cross sections,
leading to a re-calculation of new stellar opacities (Seaton \etal
1994). The OP work, based on the non-relativistic formulation of
the close coupling approximation using the R-matrix method (Seaton 1987,
Berrington \etal 1987), was carried out in LS coupling, neglecting
relativistic fine structure that is not crucial in the calculation
of mean plasma opacities. Also, collisional processes were not
considered under the OP. The IP collaboration seeks to address both of
these factors, and with particular reference to iron and iron-peak elements.
The collaboration involves members from six countries: Canada, France,
Germany, UK, US, and Venezuela.

 The relativistic extension of the R-matrix method is
based on the Breit-Pauli approximation (Berrington \etal 1995). Collisional and radiative processes may both be
considered. However, the computational requirements for the Breit-Pauli
R-matrix (hereafter BPRM) calculations can be orders of magnitude more
intensive than non-relativistic calculations. Nonetheless, a large body
of atomic data has been obtained and published in a continuing series
under the title {\bf Atomic Data from the Iron Project} in {\it Astronomy
and Astrophysics Supplement Series}, with 43 publications at present.
A list of the IP publications and related information may be
obtained from the author's Website: www.astronomy.ohio-state.edu/~pradhan.

 The earlier phases of
the Iron Project dealt with (A) fine structure transitions among low-lying
levels of the ground configuration of interest in Infrared (IR) astronomy,
particularly the observations from the Infrared Space Observatory, 
 and (B) excitation of the large number of levels in multiply ionized
iron ions (with n = 2,3 open shell electrons, i.e. Fe~VII -- Fe~XXIV)
of interest in the UV and EUV, particularly for the Solar and Heliospheric
Observatory (SOHO), the Extreme Ultraviolet Explorer (EUVE), and Far Ultraviolet
Spectroscopic Explorer (FUSE). In addition, the IP data for the low ionization 
stages of iron (Fe~I -- Fe~VI) is of particular interest in the analysis
of optical and IR observations from ground based observatories.
  In the present review, we describe the IP work within the context of
applications to X-ray spectroscopy, where ongoing calculations on 
collisional and radiative data for H-like Fe~XXVI, He-like Fe~XXV, and 
Ne-like Fe~XVII are of special interest. 

 The sections of this review are organised as follows: 1. Theoretical, 2. collisional,
3. radiative, 4. collisional-radiative modeling of X-ray spectra, 5.
atomic data, and 6. Discussion and conclusion.

\section{The Close Coupling approximation and the Breit-Pauli R-matrix Method}

In the close coupling (CC) approximation
 the total electron {+} ion wave function may be represented as
\begin{equation}
\Psi = A \sum_{i=1}^{NF} \psi_{i}\theta_{i} + \sum_{j=1} C_{j} \Phi_{J},
\end{equation}
where $\psi_{i}$ is a target ion wave function in a specific state
$S_{i}$
$L_{i}$ and $\theta_{i}$ is the wave function for the free electron in a
channel labeled as $S_{i}L_{i}k_{i}^{2}\ell_{i}(SL\pi)$, $k_{i}^{2}$
being its
incident kinetic energy relative to $E(S_{i}L_{i})$ and $\ell_{i}$ its
orbital
angular momentum.  The total number of free channels is $NF$ (``open''
or
``closed'' according to whether $k_{i}^{2}$ $<$ or $>$ $E(S_{i}L_{i})$.
$A$ is
the antisymmetrization operator for all $N+1$ electron bound states,
with
$C_{j}$ as variational coefficients.  The second sum in Eq. (1)
represents
short-range correlation effects and orthogonality constraints between
the
continuum electron and the one-electron orbitals in the target. 

 The target levels included in the first sum on the RHS of Eq.
(1) are coupled; their number limits the scope of the CC calculations.
Resonances arise naturally when the incident electron energies excite 
some levels, but not higher ones, resulting in a coupling between ``closed" and
``open" channels, i.e. between free and (quasi)bound wavefunctions.
The  R-matrix method is the most efficient means of solving the CC
equations and resolution of resonance profiles (see reviews by K.A.
Berrington and M.A. Bautista). The relativistic CC approximation may be
implemented using the Breit-Pauli Hamiltonian.

 Both the continumm wavefunctions at E $>$ 0 for the (e +ion)
system, and bound state wavefunctions may be calculated. Collision
strengths are obtained from the continuum (scattering) wavefunctions,
and radiative transition matrix elements from the continuum and the bound
wavefunctions that yield transition probabilities
and photoionization and (e~+~ion) photo-recombination cross sections 
(see the review by S.N. Nahar).

Recent IP calculations for the n~=~3 open shell ions include up to
100 or more coupled fine structure levels. Computational requirements are
for such radiative and collisional calculations may be of the order of
1000 CPU hours even on the most powerful supercomputers.

\section{Electron Impact Excitation} 

 Collision strengths and maxwellian averaged rate coefficients
have been or are being calculated for all ions of iron. While some of
the most
difficult cases, with up to 100 coupled fine strcture levels from n = 3
open shell configurations in Fe~VII -- Fe~XVII, are still in progress,
most other ionization stages have been completed. In particular fine
structure collision strengths and rates have been computes for thousands
of transitions in Fe~II -- Fe~VI. For a list of papers see ``Iron Project" 
on www.astronomy.ohio-state.edu/~pradhan. 

 Work on K-shell and L-shell collisional excitations, begining
with the H-like and the He-like ions will be continued under the new RMaX
project, which is part of the IP and is focused on X-ray spectroscopy.
Work is in progress on He-like Fe~XXV (Mendoza \etal) and Ne-like 
Fe~XVII. Fig. 1 presents the collision strength for
a transition in Fe~XVII from the new 89-level BPRM calculation
including the n = 4 complex (Chen and Pradhan 2000). The extensive
resonance structure is due to the large number of coupled thresholds
following L-shell excitation.

\begin{figure}
\centering
\psfig{figure=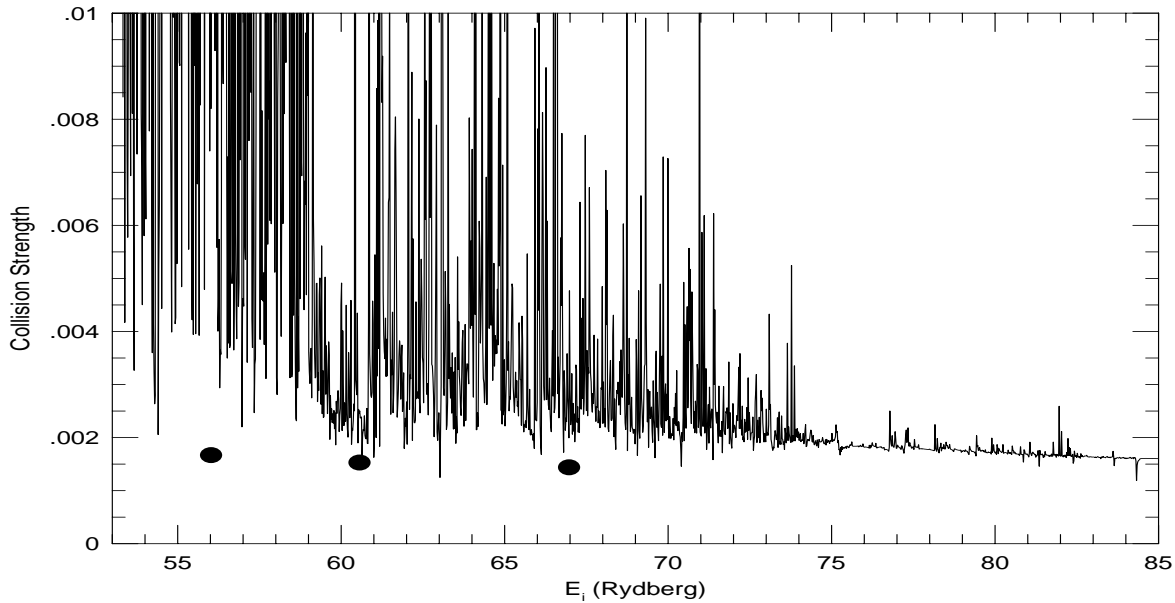,height=10.0cm,width=18.0cm}
\caption{The BPRM collision strength $\Omega (2p^6 \ ^1S_0
\longrightarrow 2p^5 \ 3s \ ^3P_2)$ (Chen and Pradhan 2000); 
the relativistic distorted wave
values are denoted as filled circles (H.L. Zhang, in Pradhan and Zhang
2000)}
\end{figure}

\section{Radiative transition probabilities}

 There are two sets of IP calculations: (i) with atomic structure 
codes CIV3 (Hibbert 1973) and SUPERSTRUCTURE (Eissner \etal 1974), and
(ii) BPRM calculations. Of particular interest to X-ray work are the
recent BPRM calcultions for 2,579 dipole (E1) oscillator strengths for 
 Fe~XXV , and 802 transitions in Fe~XXIV (Nahar and Pradhan 1999),
extending the available datasets for these ions by more than an order of
magnitude. Also, these data are shown to be highly accurate, 1 -- 10\%.

\section{Collisional-Radiative model for He-like ions: X-ray emission 
from Fe~XXV}

 Emission from He-like ions provides the most valuable X-ray spectral 
diagnostics for the temperature, density, ionization state, and other
conditions in the source (Gabriel 1972, Mewe and Schrijver
1981, Pradhan 1982).
 The K$\alpha$ complex of He-like ions consists of the principal 
lines from the allowed (w), intersystem (x,y), and
the forbidden (z) transitions $1^1S \longleftarrow 2(^1P^o, \ ^3P^o_2,
\ ^3P^o_1, \ ^3S_1$ respectively. (These are also referred as the R,I,F
lines, where the I is the sum (x+y); we employ the former notation). 
Two main line ratios are particularly useful, i.e.

\be \ R = \frac{z}{x+y} \ , \ee and
\be \ G = \frac{x+y+z}{w} \ . \ee

 R is the ratio of forbidden to intersystem lines and is sensitive to 
electron density N$_e$ since the forbidden
line z may be collisionally quenched at high densities. G is
the ratio of the triplet-multiplicity lines to the `resonance' line, and
is sensitive to (i) electron temperature, and (ii) ionization balance.
Condition (ii) results because recombination-cascades from
H-like ions preferentially populate the triplet levels, enhancing the z
line intensity in particular (the level $2(^3S_1)$ is like the `ground' 
level for the triplet levels). Inner-shell ionization of Li-like 
ions may also populate the $2(^3S_1)$ level ($1s^2
\ 2s \longrightarrow 1s2s + e)$ enhancing the z line. 
The line ratio G is therefore a sensitive indicator of the
ionization state and the temperature of the plasma during ionization,
recombination, or in coronal equilibrium.

 For Fe~XXV the X-ray lines {w,x,y,z} are at $\lambda\lambda$ 1.8505,
1.8554, 1.8595, 1.8682 $\AA$, or 6.700, 6.682, 6.668, 6.637 keV, respectively.
A collisional-radiative model (Oelgoetz and Pradhan, in progress) 
including electron impact ionization,
recombination, excitation, and radiative cascades is used to compute
these line intensties using rates given by Mewe and Schrijver (1978), 
Bely-Dubau \etal (1982), and Pradhan (1985a). New unified electron-ion
recombination rates (total and
level-specific) are being calculated by S.N. Nahar and
collaborators, and electron excitation rates are being recalculated
by C. Mendoza and collaborators; these will be employed in a more
accurate model of X-ray emission from He-like ions.

Fig. 2 shows illustrative results for doppler broadened line 
profiles under different plasma conditions (normalized to I(w) = 1). 
All are at $N_e = 10^{10} \ cm^{-3} << N_c$, so that the R dependence is 
only on T$_e$. Figs.  2(a) and 2(b) are in coronal equilibrium, but
differing widely in T$_e$, $10^7 -- 10^8$K, as reflected in the broader
profiles for the latter case. The ratios R and G show a significant
(though not large) temperature
dependence in this range. The 
ionization fractions Fe~XXIV/FeXXV and Fe~XXVI/FeXXV for the two cases
are such that the Li-like iron dominates at $10^7$K and the H-like at
$10^8$K. Figs. 2(a) and (b) illustrate a general property of the He-like
line ratios: {\it G $\approx 1$ in cororal equilibrium} (for other
He-like ions it may vary by 10-20\%).

 On the other hand, the situation is quite different when the plasma is
out of equilibrium. In particular, it is known that 
the forbidden line z is extremely sensitive to
the ionization state since it is predominantly populated via
recombination-cascades (Pradhan 1985b). Fig. 2(c) illustrates a case where
recombinations are suppressed, and the plasma is at $T_e = 10^8$ K. The
total G value is now only a third of its coronal value, with the z/w
ratio being considerably lower. Although the new recombination and
excitation rates may change the number somewhat, it is seen that 
{\it G $\approx 0.37$ is a lower limit on an ionization dominated plasma}.  

A reverse situation occurs in a recombination dominated plasma. It is
known from tomakak studies (Kallne \etal 1984, Pradhan 1985b) that the z/w ratio, and
hence G, increases practically without limit, as $T_e$ decreases much below the
coronal temperature of maximum abundance. {\it $G >> 1$ observed values
imply a recombination dominated source}. However, the z/w ratio may also
be enhanced by inner-shell ionization through the Li-like state. More
detailed calculations are needed to distinguish precisely between the
two cases, and to constrain the temperature and ionization fractions.

 Di-electronic satellite intensities (Gabriel 1972) 
 may also be computed using BPRM data for the autoionization and
radiative rates of the satellite levels from recombination of e~+~FeXXV
$\longrightarrow$ Fe~XXIV (Pradhan and Zhang 1997). This work is in
progress.

\begin{figure}
\centering
\psfig{figure=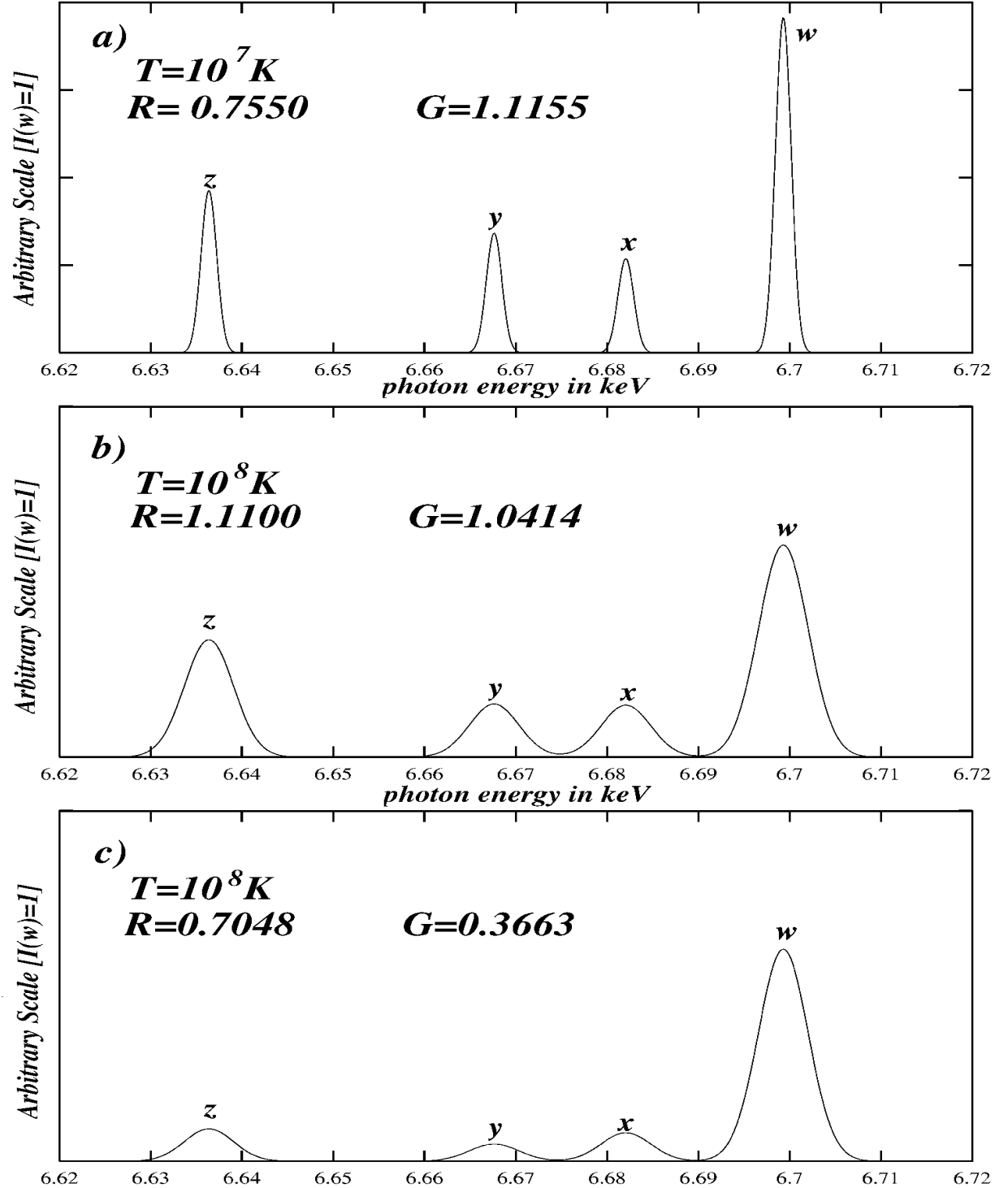,height=20.0cm,width=18.0cm}
\caption{X-ray spectra of Fe~XXV (Oelgoetz and Pradhan 2000). 
The principal lines w,x,y,z and the line ratios R and G are computed at plasma 
electron temperatues shown.
The lines are Doppler broadened. I(w) is normalized to unity.}
\end{figure}

\section{Atomic Data}

 The atomic data from the OP/IP is available from the Astronomy and
Astrophysics library at CDS, France (Cunto \etal 1993). The data is also
available from a Website at NASA GSFC linked to the author's Website
(www.astronomy.ohio-state.edu/~pradhan).

 A general review of the methods and data, 
(``Electron Collisions with Atomic Ions - Excitation",
Pradhan and Zhang 2000) is available from the author's website. The
review contains an evaluated compilation of theoretical data sources
for the period 1992-1999, as a follow-up of a similar review of all data
sources up to 1992 by Pradhan and Gallagher (1992) -- a total of over
1,500 data sources with accuracy assessment. Also contained are
data tables for many Fe ions, and a recommended data table of effective
collision strengths and A-values for radiative-collisional models for
ions of interest in nebular plasmas.

 The collisional data from the IP is being archived in a new database
called TIPBASE, complementary to the radiative database from the OP,
TOPBASE (see the review by C. Mendoza).

\section{Discussion and Conclusion}

 An overview of the work under the Iron Project collaboration was
presented. Its special relevance to X-ray astronomy was pointed out since the
IP, and related work, primarily aims to study the dominant atomic processes 
in plasmas, and to compute extensive and accurate set of atomic 
data for electron impact excitation, photoionization, recombination, and 
transition probabilities of iron and iron-peak elements.
 The importance of coupled-channel calculations was emphasized, in
particular the role of autoionizing resonances in atomic phenomena. 
(A new project RMaX, a part of IP focused on X-ray spectroscopy, is
described by K.A. Berrington in this review). 

 During the discussion, a question was raised regarding the resonances
in Fe~XVII collision strengths (e.g. Fig. 1), and it was mentioned that new 
experimental measurments appear not to show the expected rapid variations in
cross sections. A possible explanation may be that there are numerous 
narrow resonances in the entire near-threshold region, without a clearly
discernible background or energy gap. The measured cross sections are 
averages over the resonances corresponding to the experimental beam-width. 
These averaged cross sections themselves may not exhibit sharp variations,
unlike more highly charged He-like ions where the 
the non-resonant background and the resonance complexes are well
separated in energy (e.g. He-like Ti~XXI, Zhang and Pradhan 1993).


\def\amp{{\it Adv. At. Molec. Phys.}\ }
\def\apj{{\it Astrophys. J.}\ }
\def\apjs{{\it Astrophys. J. Suppl. Ser.}\ }
\def\apjl{{\it Astrophys. J. (Letters)}\ }
\def\aj{{\it Astron. J.}\ }
\def\aa{{\it Astron. Astrophys.}\ }
\def\aasup{{\it Astron. Astrophys. Suppl.}\ }
\def\adndt{{\it At. Data Nucl. Data Tables}\ }
\def\cpc{{\it Comput. Phys. Commun.}\ }
\def\jqsrt{{\it J. Quant. Spectrosc. Radiat. Transfer}\ }
\def\jpb{{\it Journal Of Physics B}\ }
\def\pasp{{\it Pub. Astron. Soc. Pacific}\ }
\def\mn{{\it Mon. Not. R. astr. Soc.}\ }
\def\pra{{\it Physical Review A}\ }
\def\prl{{\it Physical Review Letters}\ }
\def\zpds{{\it Z. Phys. D Suppl.}\ }
\def\adndt{Atomic Data And Nuclear Data Tables}

\section{References}
\parindent=0pt

Bely-Dubau, F. Dubau, J., Faucher, P. and Gabriel, A.H. 1982, \mn,
198 239\\
Berrington, K.A., Burke, P.G., Butler, K., Seaton, M.J.,
Storey, P.J., Taylor, K.T., \& Yan, Yu. 1987, \jpb 20, 6379\\
Berrington K.A., Eissner W.B., Norrington P.H.,
1995, \cpc 92, 290\\
Cunto,W.C., Mendoza,C., Ochsenbein,F. and Zeippen, C.J., 1993, \aa
 275, L5\\
Eissner W, Jones M and Nussbaumer H 1974 \cpc 8 270
Gabriel, A.H., \mn 1972, 160, 99\\
Hibbert A., 1975, \cpc 9, 141\\
Hummer, D.G., Berrington, K.A., Eissner, W., Pradhan,
A.K., Saraph, H.E., \& Tully, J.A. 1993, Astron. Astrophys. 279, 298\\
Kallne, E, Kallne, J., Dalgarno, A., Marmar, E.S., Rice, J.E. and
Pradhan, A.K. 1984, \prl 52, 2245\\
Mewe, R. and Schrijver, J. 1978, \aa, 65, 99\\
Nahar, S.N. and Pradhan, A.K. 1999, \aasup 135, 347\\
Pradhan, A.K. 1982 \apj, 263, 477\\
Pradhan, A.K. 1985a \apjs, 59, 183\\
Pradhan, A.K. 1985b \apj, 288, 824\\
Pradhan, A.K. and Gallagher, J.W. 1992, \adndt, 52, 227\\
Pradhan, A.K. and Zhang, H.L. 1997, \jpb, 30, L571\\
Pradhan, A.K. and Zhang, H.L. 2000, {\it ``Electron Collisions with
Atomic Ions"}, In LAND\"{O}LT-BORNSTEIN Volume {\it ``Atomic Collisions"},
Ed. Y. Itikawa, Springer-Verlag (in press).\\
The Opacity Project Team, {\it The Opacity Project}, Vol.1, 1995, 
Institute of Physics Publishing, U.K.\\ 
Seaton, M.J. 1987, \jpb 20, 6363\\
Seaton, M.J., Yu, Y., Mihalas, D. and Pradhan, A.K. 1994, \mn, 266, 805\\
Zhang H.L. and Pradhan A.K. 1995, \pra, 52, 3366

\end{document}